\documentclass[12pt]{article}
 
\usepackage{latexsym}
\usepackage{graphics}
\newcommand{\be}{\begin{equation}}
\newcommand{\ee}{\end{equation}}
\newcommand{\ben}{\begin{eqnarray}}
\newcommand{\een}{\end{eqnarray}}

\newcommand{\bea}{\begin{eqnarray}}
\newcommand{\eea}{\end{eqnarray}}

\begin{document}

\setlength{\baselineskip}{19pt}
\title{
\normalsize
\mbox{ }\hspace{\fill}
\begin{minipage}{7cm}
UPR-939-T\\
{\tt hep-ph/0105010}{\hfill}
\end{minipage}\\[5ex]
{\large\bf  From a Thick to  a Thin Supergravity Domain Wall 
 \\[1ex]}}
\author{Francisco A. Brito,  Mirjam Cveti\v c
and  SangChul Yoon\\
{\it Department  of Physics
and Astronomy},\\
{\it University of Pennsylvania,
Philadelphia PA 19104-6396, USA}\\
}

\maketitle  

\thispagestyle{empty}

\begin{abstract}
Within D=4 N=1  supergravity theory  we obtain an
effective theory of the  thin wall
limit for a flat domain wall configuration, interpolating
between isolated supersymmetric
 extrema of the  matter potential. We focus on the $Z_2$ symmetric flat
wall   and derive the supersymmetric  effective action
both in the bulk and on the wall; in the thin wall limit the
 scalar field, forming the wall, is frozen,
  and provides the delta function
domain wall source, while
 its fermionic partner decouples due 
to its large mass. 	In addition, the interaction between the
gravitational supermultiplet and the interaction of the matter multiplet
decouples on the wall. 
 While the results are explicitly
demonstrated  within
 $N=1$ $D=4$ supergravity, we also provide a generalization of the result
to
$D$-dimensions.

\end{abstract}
\newpage
 
\section{Introduction}
Domain wall configurations are of
special interest in gravity theory,
since
due to its minimal co-dimension, they 
are expected to exhibit strong and
nontrivial gravitational effects.
(See e.g., \cite{cv2} for a  review and references  therein.) 
In particular, flat domain wall configurations are of special
interest, since (within $D=4$ $N=1$ supergravity) they were shown
\cite{cvr} to correspond to  supersymmetric (BPS) configurations,
interpolating between supersymmetric
isolated extrema  of the
matter potential.   

In particular the thin wall limit of the $Z_2$ symmetric flat domain wall
configuration, interpolating between negative cosmological constant
isolated extrema in $D=5$ dimensions, has  been a focus of intense study
since it
allows for trapping of gravity \cite{rs2} on the wall.  
In the original version the fixed relationship between the energy density
of the wall and the cosmological constant of the bulk was viewed as a
fine-tuning condition.  It was further pointed out that this condition is
related to the BPS mass-density formula for a $Z_2$ symmetry BPS limit
\cite{cvr,cvb,skt}.

The purpose of this paper is to explore in detail the thin wall limit of a
flat $Z_2$  symmetric domain wall within $D=4$ $N=1$ supergravity theory. In
particular, we will determine the
effective action for such a flat wall configuration as one approaches
the thin wall limit. We shall identify the bulk and the domain wall
part of the action both for the bosonic and fermionic sector of the theory
and
discuss the  properties of supersymmetry transformations in this 
limit.  Details are given in Section 2.

In the second part we shall generalize the procedure to an
``effective'' supergravity theory in D-dimensions, following  Ref.\cite{gbl}: in
the gravity
supermultiplet the turned on
fields  are chosen to be  the graviton and  gravitino, and in the matter
supermultiplet  the scalar field (forming a classical
kink solution) and the spin-one half partner field are turned on. The details  of
deriving
the thin wall limit effective Lagrangian are given in Section 3.
Conclusions are given in Section 4.



\section{BPS Thin Wall Limit Effective Lagrangian within N=1 D=4
Supergravity}

The starting point is 
$N=1$ $D=4$ Lagrangian with one  neutral chiral supermultiplet whose
superpotential allows for a formation of the $Z_2$ symmetric kink
solution,
forming a BPS domain wall,  first discussed in \cite{cvr}.
We first recapitulate these results. Then we turn to the derivation of the
thin domain wall and further derive the effective supersymmetric action
associated with this thin wall  BPS domain wall configuration.


The $N=1$  $D=4$  supergravity Lagrangian 
 with one neutral chiral supermultiplet
  is of the form (\cite{wbg}):
\bea
\label{yeq00}
 e^{-1} \mathcal{L} & = & - \frac{1}{2}{R} - | \partial_{z}
 \phi |^{2} - i \bar{\chi} \bar{\sigma}^{m} \mathcal{D}_{m} \chi +
 \epsilon^{klmn} \bar{\psi}_{k}
 \bar{\sigma}_{l} \tilde{\mathcal{D}}_{m} \psi_{n}  \nonumber \\
 & & - \frac{\sqrt{2}}{2} \partial_{z} \bar{\phi} \chi \sigma^{m}
\bar{\sigma}^{z}
 \psi_{m} - \frac{\sqrt{2}}{2} \partial_{z} \phi \bar{\chi}
\bar{\sigma}^{m} \sigma^{z}
 \bar{\psi}_{m} \nonumber \\
 & & + \frac{1}{4} [i \epsilon^{klmn} \psi_{k} \sigma_{l} \bar{\psi}_{m}
 + \psi_{m} \sigma^{n} \bar{\psi}^{m}]
 \chi \sigma_{n} \bar{\chi} - \frac{1}{8} \chi \chi \bar{\chi} \bar{\chi}
\nonumber \\
 & & - e^{\frac{K}{2}} \{ \overline{W} \psi_{a} \sigma^{ab} \psi_{b}
 + W \bar{\psi}_{a} \bar{\sigma}^{ab} \bar{\psi}_{b} + i
\frac{\sqrt{2}}{2}
 D_{\phi} W \chi \sigma^{a} \bar{\psi}_{a} + i \frac{\sqrt{2}}{2}
D_{\bar{\phi}} \overline{W}
 \bar{\chi} \bar{\sigma}^{a} \psi_{a} \nonumber \\
 & & + \frac{1}{2} \mathcal{D}_{\phi} D_{\phi} W \chi \chi
 + \frac{1}{2} \mathcal{D}_{\bar{\phi}} D_{\bar{\phi}} \overline{W}
\bar{\chi} \bar{\chi} \} \nonumber \\
 & & - e^{K} \{ | D_{\phi} W |^{2} - 3 | W |^{2} \}\, .
\eea\
 Here  ($e^a_m$, $\psi_m$) denotes the supergravity
multiplet  components   and  ($\phi$, $\chi$) the  chiral superfield
components. We
have set $M_{pl}=1$. (For  notation and conventions see
\cite{wbg}.)

The covariant derivatives are defined  in the standard form:
\bea && \mathcal{D}_{m} \chi = \partial_{m} \chi +\chi
\omega_{m}-\frac{1}{4} (K_{\phi} \partial_{m} \phi -
K_{\bar{\phi}}\partial_{m} \bar{\phi})\chi \, ,\nonumber\\
&& \tilde{\mathcal{D}}_{m} \psi_{n}= \partial_{m} \psi_{n}+\psi_{n}
\omega_{m}+\frac{1}{4}(K_{\phi} \partial_{m}\phi-K_{\bar{\phi}}
\partial_{m}\bar{\phi})\psi_{n}\, , \nonumber\\
&& D_{\phi}W=W_{\phi}+K_{\phi} W, \qquad W_{\phi}=\frac{\partial
W}{\partial \phi }, 
\qquad K_{\phi}=\frac{\partial K}{\partial \phi}\, ,\nonumber\\
&&\mathcal{D}_{\phi}
D_{\phi}W=W_{\phi\phi}+K_{\phi\phi}W+2K_{\phi}D_{\phi}W-K_{\phi}K_{\phi}W\,
.
\eea

The specific form of the superpotential $W$ and the minimal K\"ahler
potential
$K$ is chosen to be of the form:
\be
W=\lambda\left(\frac{\phi^3}{3}-\eta^2\phi\right), \ \ K=\phi \bar{\phi} . 
\ee
This choice 
allows for a  BPS domain wall configuration Ansatz interpolating between
two
isolated supersymmetric extrema of the potential determined by the
solution:
$D_{\bar\phi} \overline{W}=0$:
\bea
D_{\bar{\phi}}\overline{W}&=&\lambda
 \left[\frac{1}{3}|\phi|^{2}
(\bar{\phi}^{2}-3\eta^{2})+\bar{\phi}^{2}-\eta^{2}\right]=0\, ,
 \nonumber\\
\phi_{\pm}&=&\pm
\left[\frac{3\eta^{2}-3+\sqrt{9\eta^{4}-6\eta^{2}+9}}{2}\right]^{\frac{1}{2}}
\,
. 
\label{sv}\eea  
The  domain wall Ansatz for the
 scalar component of the 
 chiral supermultiplet and the metric are respectively 
  ${\rm Re}\phi =\varphi(z)$ and:
\be ds^2=e^{2A(z)}\eta_{\mu\nu}dx^{\mu}dx^{\nu}+dz^2\, , \qquad\mu,\nu=0,1,2\, .
\ee
We employ the Killing spinor equations: $\delta\chi=0$ and
$\delta\psi_m$=0.
Here  the gravitino transformation takes the form:
\be \delta\psi_{m}=2\mathcal{D}_{m}\epsilon+i
e^{\frac{K}{2}}W \sigma_{m} \bar{\epsilon}\, , 
\label{eq.8a}\ee 
and the transformation on the fermionic partner is of the form:
\ben
\label{eq.8b}
\delta\chi=i\sqrt{2}\sigma^m\bar{\epsilon}\partial_m\phi-\sqrt{2}e^{K/2}
K^{\phi\bar{\phi}}D_{\bar{\phi}}
\overline{W}\epsilon\, .
\een
 
For BPS supergravity domain walls, first
obtained in \cite{cvr},   the equation for
  $\varphi(z)$  and $A(z)$ arise as solutions of the Killing spinor
equations
  which take the following form:
\be
\label{yeq0}
\partial_z\phi(z)=ie^{i\theta}e^{\frac{K}{2}}K^{\phi\bar{\phi}}
D_{\bar{\phi}}\overline{W}\, ,
\, \ee
with  the  two component spinor
 $\epsilon = {\epsilon_1
\choose \epsilon_2}$ satisfying $\bar{\sigma}^z\bar{\epsilon}=e^{-i\theta}\epsilon$,
 where the  phase $\theta$ satisfies the following ``geodesic'' equation:
 \be
\partial_{z} \theta =-Im(K_{\phi}\partial_{z}\phi)\simeq 0\, ,
\label{phase}\ee
and is satisfied for the Ansatz ${\rm Re}\phi(z) = \varphi(z)$, $\theta
=\frac{\pi}{2}$.
The warp factor satisfies the following equation (arising from:
$\delta\psi_t=0$): 
\be
\label{yeq1a}
\partial_z A(z)=e^{A(z)}e^{\frac{K}{2}}W\, .
\ee 
 
%
Note again,  that while $\phi$ is a complex  field, only its real component ${\rm
Re}\phi \equiv \varphi(z)$ is turned on as  a kink solution. This Ansatz is
indeed
compatible with the Killing spinor equations, thus preserving $1\over 2$
of the
original supersymmetry. (See also  \cite{cvr} for  further details.)



The solution was solved numerically in \cite{cvr} and has  the following
properties: the scalar field $\varphi(z)$ is a typical kink approaching
exponentially fast the asymptotic values $\varphi(z\to\pm
\infty)=\phi_{\pm}$,  given by (\ref{sv}). The warp factor of the metric
approaches exponentially fast the Cauchy horizon of the anti-deSitter
(AdS) space
times of the supersymmetric ground states on either side of the wall, i.e.
$e^{2A(z)}\sim{4}/{\sigma^2 z^2}$ as $z\to \pm \infty$, 
where $\sigma=(4/3)\lambda\eta^3.$

We now turn to the thin wall limit by taking
$\lambda \rightarrow \infty$, $\eta\rightarrow 0$, and $\lambda\eta^3\to$
fixed.
We shall first focus on the bosonic part of the effective Lagrangian in
the thin
wall  background.
In this  limit,   eq.(\ref{sv}) implies that $\phi_{\pm}
\simeq
\pm \eta$. In this case  the kink  solution of (\ref{yeq0}) assumes the form:
\be 
\label{yeq6}
\varphi(z)\simeq \eta \tanh{\lambda\eta z} \, ,
\ee
i.e. it  approaches the  step function 
$\varphi(z)\simeq\eta\,{\rm sign}\,z$, whose 
width goes to zero. 
We introduce some useful ``identities'':
\ben
\label{eq6ab}
& &W\simeq-\frac{2}{3}\lambda\eta^3\,{\rm sign}\,z,\ \ D_\phi
W\simeq-2\eta\delta(z),\ \ 
\partial_z\phi\simeq2\eta\delta(z)\, ,\nonumber\\
& &\left[\partial_z\phi(z)\right]^2=\left[D_\phi
W\right]^2\simeq\sigma\delta(z),
\ \ {\mathcal D}_\phi D_\phi W\simeq2\lambda\eta\,{\rm sign}\,z, \ \
|W|^2\simeq
\frac{4\lambda^2\eta^6}{9}\, . 
\een
(Of course the expressions with the  delta functions  make sense 
only under integration.)
We can now see easily that the solution (\ref{yeq6}) satisfies  the thin
wall limit of 
the equation (\ref{yeq0}). It is also instructive to see that in this
limit the equation 
(\ref{yeq1a}) for the metric becomes simply
\ben
\label{eq6abc}
\partial_z [e^{-A(z)}]\simeq -W\simeq\frac{2}{3}\lambda\eta^3{\rm
sign}\,z=\frac{1}{2}\sigma\,{\rm sign}\,z\, ,
\een
which has the ``exact" solution $e^{2A(z)}\simeq4/\sigma^2 z^2$. Note that the warp factor 
here agree with the asymptotic behavior discussed earlier. 
 
In the thin wall limit the supersymmetric
transformation (\ref{eq.8a})  has  the factor of the second term in  the
transformation,   $e^{\frac{K}{2}}W \sigma_{m}$,  approaching the step
function with the finite width: 
\ben\label{eq6.b}
\delta\psi_{m}\simeq2\mathcal{D}_{m}\epsilon-i\left(\frac{1}{2}\sigma\,{\rm
sign}\,z\right)
\sigma_{m} 
\bar{\epsilon}\, . 
\een
Note that since $W$ changes the sign, i.e. it is
approximately a step function with a {\it finite} width,  the effective 
gravitino mass term in the Lagrangian takes the form:
\ben
\label{eq6.c}
{\cal L}_{\psi^2}=\frac{1}{2}\sigma\,{\rm
sign}\,z\,(\psi_{a}\sigma^{ab}\psi_{b}+\bar{\psi}_{a}
\bar{\sigma}^{ab}\bar{\psi}_{b})\, , 
\een
i.e. on the either side of the wall it  has  an opposite sign.
Note that this  result  is   
in agreement with the  modified  supergravity transformation employed
by \cite{flp}. 

Similarly,  transformation (\ref{eq.8b}) takes the form:
\ben
\delta\chi\simeq 2\sqrt{2}\eta(i\sigma^z\bar{\epsilon}+\epsilon)\delta(z)\, ,
\een
i.e. there is no contribution from the bulk and on the wall the prefactor goes
to zero.



We shall now  write down  the structure of the 
 thin wall effective  action. For the
bosonic sector,  using the results (\ref{eq6ab}), 
the action becomes:
\bea
S^{eff}=\int_{bulk} d^{4}x e\left[-\frac{{R}}{2}+\Lambda\right]+
T\int_{brane} d^{4}x 
 e\delta(z)\, . 
\eea
The anti-deSitter (AdS) cosmological constant $\Lambda=(3/4)\sigma^2$  arises
 from the last term in
the scalar potential and the $\delta$-function brane tension $T=-2\sigma$
effectively arises from the
kinetic energy of $\phi$ and the first term of the supergravity
scalar potential in (\ref{yeq00}). The tension and the cosmological constant 
 are related  as 
$\Lambda=(3/16)T^2$. This
action is  precisely
 the action of the type  used in \cite{rs2}; therefore
the fine tuning
between the AdS cosmological constant and the brane tension is, within our
framework, a consequence of supersymmetry. 
(Recall that $\sigma=(4/3)\lambda\eta^3$).

For the fermionic sector, the action (\ref{yeq00}) becomes: 
\bea
S_{bulk}^{eff}&=&\int
d^{4}e\left[-\frac{1}{2}{R}-i\bar{\chi}\bar{\sigma}^{m}\mathcal{D}_{m}
\chi+\epsilon^{klmn}\bar{\psi}_{k}\bar{\sigma}_{l}\tilde{\mathcal{D}}_{m}\psi_{n}\right.\nonumber\\
&+&\frac{1}{4}(i\epsilon^{klmn}\psi_{k}\sigma_{l}\bar{\psi}_{m}+\psi_{m}\sigma^{n}\bar{\psi}^{m}
)
\chi\sigma_{n}\bar{\chi}-\frac{1}{8}\chi\chi\bar{\chi}\bar{\chi}\nonumber\\
&+&\frac{1}{2}\sigma\;\mbox{sign}\;{z}(\psi_{a}\sigma^{ab}\psi_{b}+\bar{\psi}_{a}\bar{\sigma}^{ab}
\bar{\psi}_{b})
-\frac{3}{4}\frac{\sigma}{\eta^2}\;\mbox{sign}\;{z}(\chi\chi+\bar{\chi}\bar{\chi})\nonumber\\
&+&\left.\frac{3}{4}\sigma^2\right]\, , \nonumber\\
S_{brane}^{eff}&=&\int d^{4} e\delta(z)[-2\sigma-\sqrt{2}\eta
(\chi\sigma^{m}\bar{\sigma}^{z}\psi_{m}+\bar{\chi}\bar{\sigma}^{m}\sigma^{z}\bar{\psi}_{m})\nonumber\\
&+&i\sqrt{2}\eta(\chi\sigma^{a}\bar{\psi}_{a}+\bar{\chi}\bar{\sigma}^{a}\psi_{a})]
\, .
\eea 
 The limit $\eta \to 0$, implies that the fermionic fields 
 $\chi$,$\psi_m$  do not contribute  to the 
 brane part  of the  effective action.  In addition,  in the bulk, 
  the fermionic field
$\chi$ has a  large mass $(3/2)(\sigma/\eta^2)\to \infty$ and  thus 
decouples from the
theory. 
This loss of  $\chi$  fermionic degrees of freedom 
parallels the fact that the kink, associated with the 
 bosonic field 
 partner, has been
frozen  to be ``infinitely stiff" step function.

\section{Effective action from a supergravity model in D-dimensions}

In this section we  employ  an effective supergravity action in 
D-dimensions
derived in 
\cite{gbl} where  most of the attention was focused on
D=odd dimensions.   The Lagrangian contains the graviton and
gravitino as the only fields  turned on in the supergravity multiplet, 
and the real scalar and a spin 1/2  fermionic 
partner as the fields turned on in the matter supermultiplet; it can be viewed as a consistent truncation  of a supergravity
Lagrangian.  (For the sake of
simplicity we again focus on only  one  matter supermultiplet responsible for
the formation of the
wall.)
Employing  the  supersymmetry transformations
for the fermionic  fields the authors of \cite{gbl} obtained the
equations of motion for the fermionic modes which can be lifted in a
straightforward to obtain a supersymmetric Lagrangian, as described below.

The purpose of this section is to 
investigate the same issue as we did in the D=4, N=1 supergravity
 case by considering
now a consistently truncated supergravity action in D-dimensions, with
the supergravity multiplet $(e_m^a,\psi_m)$ and the scalar supermultiplet 
$(\phi,\chi)$.
We  focus on  the thin wall limit of a thick
smooth domain wall within this context.

Equations of motion for the Dirac fermion fields
coupled  to one real scalar field, a special case of the multi-field example 
studied in \cite{gbl}, are of the
form:
\ben
\label{eq5}
\Gamma^m\nabla_m\chi+M\chi+\frac{1}{2}\Gamma^m\Gamma^n\partial_n\phi\,\psi_m-W_2
\Gamma^m\psi_m=0\, ,\\
\label{eq6}
\Gamma^{mnp}\nabla_n\psi_p+W_2\Gamma^m\chi-(D-2)W\Gamma^{mn}\psi_n+\frac{1}{2}(g^{mn}-
\Gamma^{mn})\partial_n\phi\,\chi=0\, ,
\een
where 
\ben
\label{eq4} 
M&=&2(D-2)\frac{\partial^2W}{\partial\phi^2}-(D-2)W\, . 
\een
These equations were shown to be invariant under the following
supersymmetry transformations 
\cite{gbl}:
\ben
\label{eq7}
\delta\psi_m&=&\nabla_m\epsilon+W\Gamma_m\epsilon\, ,\\
\label{eq7a}
\delta\chi&=&\left(-\frac{1}{2}\Gamma^m\partial_m\phi+W_2\right)\epsilon\, ,
\een
where
\ben
\label{eq8}
\nabla_m\epsilon=\partial_m\epsilon+\frac{1}{4}\omega_m^{ab}\Gamma_{ab}\epsilon,
\qquad \Gamma_{ab}=
\frac{1}{2}[\Gamma_a,\Gamma_b]\, .
\een
In Ref. \cite{gbl} the authors use the supersymmetry transformations
(\ref{eq7})-(\ref{eq7a})
in order to construct a Nester tensor \cite{nes} and imposing the Witten
condition $\Gamma^m\delta\psi_m=0$
\cite{wit} to ensure the stability of BPS backgrounds as it is usual in
supergravity theories (and within the supergravity domain wall context first
studied in \cite{cvr}).  The preservation of supersymmetry  implies:
\ben
\label{eq2}
V(\phi)&=&4(D-2)^2\left[\left(\frac{\partial
W}{\partial\phi}\right)^2-\left(\frac{D-1}{D-2}
\right)W^2\right]\, , \\
\label{eq3}
W_2&=&(D-2)\frac{\partial W}{\partial\phi}\, .
\een

As the  next step we write down the
consistently truncated supergravity action in D-dimensions (compatible with the
above fermionic supersymmetry transformations (\ref{eq7})-(\ref{eq7a})).
Employing results \footnote{Note that one has used the fact
$\Gamma^n\Gamma^m=-\Gamma^{mn}+g^{mn}$ in the equation
(\ref{eq6}). This follows from the definitions
$\{\Gamma^m,\Gamma^n\}=2g^{mn}$ 
and $\Gamma^{mn}=
\frac{1}{2}[\Gamma^m,\Gamma^n]$.} 
(\ref{eq5})-(\ref{eq6}), (\ref{eq2})-(\ref{eq3}) and also including both
the ``Einstein-Hilbert" term 
and the kinetic part of the scalar field, yields:
\ben
\label{eq1}
S&=&\int d^Dx\,e\left[-R-g^{mn}\partial_m\phi\partial_n\phi-V(\phi)
-\bar{\psi}_m\Gamma^{mnp}\nabla_n\psi_p
-\bar{\chi}\Gamma^m\nabla_m\chi
-M\bar{\chi}\chi\right.\nonumber\\
&-&\frac{1}{2}\partial_n\phi(\bar{\psi}_m\Gamma^n\Gamma^m\chi+\bar{\chi}\Gamma^m\Gamma^n\psi_m)-
W_2(\bar{\psi}_m\Gamma^m\chi-\bar{\chi}\Gamma^m\psi_m) 
\nonumber\\
&+&\left.(D-2)W\bar{\psi}_m\Gamma^{mn}\psi_n+\mbox{ four 
fermi terms }
\right]\, . 
\een
We use the following notation:
$\Gamma^{mnp}\equiv\Gamma^{[m}\Gamma^n\Gamma^{p]}$,
$\Gamma^m=e^m_a\Gamma^a$, 
where $\Gamma^a$ are 
$2^{[D/2]}\times2^{[D/2]}$ Dirac
matrices and $e=\det{e_k^a}=|\det{g_{mn}}|^{1/2}$, with a mostly plus
signature $(-++...+)$.

\subsection{BPS domain wall and its thin limit}

The BPS domain wall background is obtained by solving Killing spinor
equations, that is, by making
(\ref{eq7})-(\ref{eq7a}) to vanish
(Note the subsequent analysis is  completely parallel to that of the
previous section.):
\ben
\label{eq9}
\partial_m\epsilon+\frac{1}{4}\omega_m^{ab}\Gamma_{ab}
\epsilon+W\Gamma_m\epsilon=0\, , \\
\label{eq10}
\left(-\frac{1}{2}\Gamma^m\partial_m\phi+W_2\right)\epsilon=0\, .
\een
The bosonic Ansatz is of the form:
the scalar field $\phi=\phi(z)$ and the
metric:
\be
\label{eq11}
ds^2=e^{2A(z)}\eta_{\mu\nu}dx^\mu dx^\nu+dz^2\, , 
\ee
where
$\mu,\nu=0,1,2,...,D-2$ are  world-volume (flat) indices and 
$z$ is the direction transverse to the wall.

With the above bosonic Ansatz  eqs. (\ref{eq9}) and (\ref{eq10})
yield
respectively:
\ben
\label{eq12}
\partial_zA=\mp2W \, , \qquad\qquad \partial_z\phi=\pm2(D-2)\frac{\partial
W}{\partial\phi}\, . 
\een
 The Killing spinors satisfy
$\epsilon=e^{\frac{1}{2}A(z)}\epsilon_\pm$,
where $\Gamma^z\epsilon_\pm=\pm\epsilon_\pm$ and 
$\Gamma_{\underline{z}}\epsilon_\pm=\pm\epsilon_\pm$. (The underlined
indices refers to the tangent frame.)

With the $Z_2$ symmetric  choice of the superpotential:
\ben
\label{e14}
W=\lambda\left(\frac{\phi^3}{3}-\eta^2\phi\right)\, ,
\een
i.e. its  critical points transform into each other under $Z_2$ symmetry (for several examples of
other superpotentials see, for instance, 
the references in \cite{bbb}). Again, our focus  will be  on the   thin wall limit
$\lambda\to\infty$, $\eta\to0$ and $\lambda\eta^3\to$ fixed.

The Bogomol'nyi equations (\ref{eq12}) for the scalar field $\phi$
can be now written as:
\ben
\label{e15}
\partial_z\phi=\pm2\lambda(D-2)(\phi^2-\eta^2)\, , 
\een 
which yields the domain wall solutions 
\ben
\label{e16}
\phi_{\pm}(z)=\mp\eta\tanh{z/\Delta}\, ,
\een
where $\Delta=1/\lambda'\eta$ specifies the domain wall width and
$\lambda'=2\lambda(D-2)$. 

In the thin wall limit we have: $\lambda\to\infty$, $\eta\to0$,
$\lambda\eta^3\to$ fixed and $\Delta\to 0$. Now the solution (\ref{e16})  takes
the form:
\ben
\label{e17}
\phi_{\pm}(z)=\mp\eta\;\mbox{sign}\;{z}\, .
\een
In addition, the superpotential in this background solution is given
by:
\ben
\label{e18}
W_+=-W_-=\left(\frac{2}{3}\lambda\eta^3\right)\;\mbox{sign}\;{z}\, ,
\een
i.e. it is a step function with a finite width. 
Here $W_{\pm}$ stand for the first and second set of
Bogomol'nyi
equations (\ref{eq12}). Regardless of the set of equations we choose
we find the very same differential equation for the metric:
\ben
\label{e19} 
\partial_z A=-\sigma\;\mbox{sign}\;{z},\qquad\qquad
\sigma=(4/3)\lambda\eta^3\, . 
\een
As we shall see $\sigma$ is proportional to the wall (brane) tension. 

The Bogomol'nyi equation (\ref{e19}) has the exact solution
$A=-\sigma|z|$.
Substituting this solution into the metric (\ref{eq11}) we find
\ben
\label{eq21}
ds^2&=&e^{-2\sigma|z|}\eta_{\mu\nu}dx^\mu dx^\nu+dz^2\, . 
\een
This is the metric used in the Randall-Sundrum scenario (in $D=5$) which is
recognized as a $Z_2$ symmetric AdS background and allows for trapping of
gravity on the domain wall.

Let us now compute the effective action for  the bosonic background. The
bosonic part of the 
action (\ref{eq1}) can be written as : 
\ben
\label{eq22}
S=\int d^Dx
e\left\{-R-(\partial_z\phi)^2-4(D-2)^2\left[\left(\frac{\partial W}
{\partial\phi}\right)^2-\left(\frac{D-1}{D-2}
\right)W^2\right]\right\}\, . 
\een
In order to derive the thin wall limit of this action, 
we employ the solution (\ref{e16}) and derive the following ``identities":
%
\ben
\label{eq23a}
& &\partial_z\phi_\pm\simeq\mp2\eta\delta(z), \ \
[\partial_z\phi_\pm(z)]^2\simeq\frac{4}{3}\lambda'
\eta^3\delta(z)=\sigma'\delta(z),\ \
\frac{\partial
W(\phi_\pm)}{\partial\phi}\simeq-\frac{\eta}{(D-2)}\delta(z),\nonumber\\
& &\left[\frac{\partial
W(\phi_\pm)}{\partial\phi}\right]^2\simeq\frac{\sigma'}{4(D-2)^2}\delta(z),\ \
[W(\phi_\pm)]^2\simeq\frac{4\lambda^2\eta^6}{9},\ 
\frac{\partial^2
W(\phi_\pm)}{\partial\phi^2}\simeq\mp2\lambda\eta\;\mbox{sign}\;{z}\, .\nonumber 
\een
We introduced new parameters $\lambda'=2\lambda(D-2)$ and
$\sigma'=2\sigma(D-2)$.  (Again, the expressions with delta functions should be
understood to make
sense only under 
integration.)
In this case the action (\ref{eq22}) takes the  form of  the following 
effective
action:
 \ben
\label{eq28}
S^{eff}=\int_{bulk} d^Dx\,e\left(-R + \Lambda\right)+T\int_{brane}
d^Dx\,e\delta(z)\, , 
\een
where 
\ben
\label{eq29}
\Lambda&=&(D-2)(D-1)\sigma^2\, , \\
\label{eq30}
T&=&-4(D-2)\sigma\, , 
\een
are the cosmological constant and the brane tension, respectively. They are
related
by 
\ben
\label{eq31}
\Lambda=\left(\frac{D-1}{D-2}\right)\frac{1}{16}T^2\, .
\een
It is instructive to consider the case $D=5$. In this case we find
$
\Lambda=12\sigma^2
$, which is exactly the Randall-Sundrum fine-tuning relation.  
(Recall  that in
\cite{rs2}
one has $\Lambda\to-\Lambda/2$, $\sigma\to k$ and Planck mass $M_{pl}\neq
1$.)

Finally, the complete effective bulk and brane action can be written as:
\ben
\label{eq34}
S_{bulk}^{eff}&=&\int d^Dx\,e\left\{-R +\Lambda
-\bar{\psi}_m\Gamma^{mnp}\nabla_n\psi_p-\bar{\chi}
\Gamma^m\nabla_m\chi\right.\nonumber
\\
&\pm&g\left(\frac{6}{\eta^2}
+1
\right)\bar{\chi}\chi
\pm\left. g\bar{\psi}_m\Gamma^{mn}\psi_n+ 
\mbox{ four fermi terms }\right\}\, , 
\een
where $g$ is the gravitino mass given by
\footnote{Note that the gravitino mass mantain 
its sign by simply ``rotating" 
the chiral fermion field with respect to
 the wall as $\psi_\pm \to \pm\Gamma^z  \psi_\pm$.}:
\ben
g=\frac{\Lambda}{2(D-1)\sigma}\;\mbox{sign}\;{z}\, , 
\een
and:
\ben
\label{eq35}
S_{brane}^{eff}&=&\int d^Dx\,e\delta(z)\left[ T \pm\eta(
\bar{\psi}_m\Gamma^z\Gamma^m\chi+\bar{\chi}\Gamma^m\Gamma^z\psi_m)\right.
\nonumber\\
&+&\left.\eta(\bar{\psi}_m\Gamma^m\chi-\bar{\chi}\Gamma^m\psi_m)
\right]\, . 
\een
 
We also see  that in the thin wall limit the supersymmetry transformations 
(\ref{eq7}) and (\ref{eq7a}) become
\ben
\label{eq35a}
\delta\psi_m&=&\nabla_m\epsilon\pm\frac{g}{(D-2)\sigma}
\Gamma_m\epsilon\, ,\\
\label{eq35b}
\delta\chi&=&(\pm\Gamma^z\epsilon-\epsilon)\eta\delta(z)\, .
\een
From eq.(\ref{eq34}) we notice that as $\eta\to 0$ (the thin
wall limit)
the mass of the fermion $\chi$ diverges because the term $6/\eta^2$ goes
to infinity. 
However, we can see clearly that in this limit the fermionic coupling in
the brane action goes
to zero as well. Thus, except for the   four-fermi terms (whose detailed
structure is not determined at the linearized level of the supersymmetry
transformations), there is no other coupling between
the fermions fields  in the theory. Unlike the  the graviton 
field, the fermion $\chi$  decouples from the theory due to its infinite mass.
 Furthermore as we can see in
(\ref{eq35b}) the supersymmetry variation $\delta\chi$
 does not have a contribution from the bulk, and the contribution from the 
 brane has a prefactor that goes to zero.

As a special case of the effective theory in $D=5$, the  term in the Lagrangian,
corresponding to the   gravitino mass
 is given by
\ben
\label{eq38}
{\cal L}_{\psi^2}^{eff}=\pm
e\left(\frac{\Lambda}{8\sigma}\;\mbox{sign}\;{z}\right)
\bar{\psi}_m\Gamma^{mn}\psi_n\, , 
\een
and the supersymmetry transformation (\ref{eq35a})  takes the form:
\ben
\label{eq39}
\delta\psi_m&=&\nabla_m\epsilon\pm\frac{\Lambda\;\mbox{sign}\;{z}}{24\sigma}
\Gamma_m\epsilon\, . 
\een
Similar equations have been considered in the reference \cite{flp} in attempting
to supersymmetrize the Randall-Sundrum action by including the term with ``sign $z$"
by hand. Here, this term appear as a natural
 consequence of the thin wall limit of the thick wall BPS solution.

\section{Conclusions}

In this paper we focused on the  structure of the supersymmetric effective
action  that is a thin wall limit of the  thick BPS domain wall solution of D=4
N=1 supergravity theory. We also explore  a generalization  to  the  
 D-dimensional,  consistently truncated supergravities.
In this limit, the kink solution for the scalar field $\phi$
provides a  background  for the  delta-function stress-energy 
 source for  the thin wall and the bulk corresponds to the 
 AdS spacetime. In this 
limit the fermionic  superpartner $\chi$  becomes infinitely heavy in the bulk
and does not have a coupling to the graviton on the brane; it thus decouples
from the theory. 
Furthermore, in  the thin wall limit there is no 
 fermionic  contribution
to the brane action.  We have also explored the structure of the supersymmetry
transformations in this limit, and arrived at the similar
supersymmetry  transformations \cite{bkp} as in \cite{flp} (see also\cite{abn}),  employed in attempts to 
supersymmetrize the Randall-Sundrum scenario; in our case these
transformations are   a natural  consequence of taking a thin wall limit for a
smooth BPS domain wall solution. 

\section{Acknowledgments}

We would like to thank Neil Lambert, Eduardo Lima, Asad Naqvi, Christopher Pope,
Gary Shiu and Kellog Stelle for helpful discussions.  
The work is supported in part by DOE
grant DE-FG02-95ER40893 and NATO grant 976951. MC would like to thank
SISSA,
Italy for hospitality during the initial stages of the work.
FAB would like to thank Department of Physics and Astronomy, University of
Pennsylvania, 
for hospitality and Conselho Nacional de Desenvolvimento Cient\'\i fico e
Tecnol\'ogico, 
CNPq, Brazil, for support.


\end{document}